\documentclass[12pt,epsf,amstex]{article}
\usepackage [dvips]{graphicx}
\usepackage{amsmath}
\usepackage{amssymb}
\usepackage{epsfig}
\usepackage{verbatim}
\usepackage{color}

\definecolor{refkey}{rgb}{0,0.5,0.5}
\definecolor{labelkey}{rgb}{0.5,0,0.5}
\definecolor{orange}{rgb}{1.0,0.75,0}

\addtocounter{secnumdepth}{1}
\def\bluew#1{{\color{blue} #1}}
\def\orangew#1{{\color{orange} #1}}

\begin{document}

\newcommand{\ee}{{\rm e}}
\newcommand{\dd}{{\rm d}}
\newcommand{\p}{\partial}
\newcommand{\phX}{\phantom{XX}}

\newcommand{\bc}{\mathbf{c}}
\newcommand{\bC}{\mathbf{C}}
\newcommand{\bex}{\boldsymbol{e}_x}
\newcommand{\bey}{\boldsymbol{e}_y}
\newcommand{\bexy}{\boldsymbol{e}_{x,y}}
\newcommand{\bq}{\mathbf{q}}
\newcommand{\br}{\mathbf{r}}
\newcommand{\bv}{\mathbf{v}}

\newcommand{\pE}{{\rm E}}
\newcommand{\pN}{{\rm N}}

\newcommand{\pEN}{\pE,\pN}
\newcommand{\pNE}{\pN,\pE}
\newcommand{\rhobar}{\overline{\rho}}
\newcommand{\rhop }{\rho^{\pE}}
\newcommand{\rhom }{\rho^{\pN}}
\newcommand{\rhopm}{\rho^{\pEN}}
\newcommand{\rhomp}{\rho^{\pNE}}
\newcommand{\rhobarp}{\overline{\rho}^{\pE}}
\newcommand{\rhobarm}{\overline{\rho}^{\pN}}
\newcommand{\rhobarpm}{\overline{\rho}^{\pEN}}
\newcommand{\etabar}{\overline{\eta}}
\newcommand{\etabarp}{\overline{\eta}^{\pE}}
\newcommand{\etabarm}{\overline{\eta}^{\pN}}
\newcommand{\etap}{\eta^{\pE}}
\newcommand{\etam}{\eta^{\pN}}
\newcommand{\etapm}{\eta^{\pEN}}
\newcommand{\drhop }{\delta\rho^{\pE}}
\newcommand{\drhom }{\delta\rho^{\pN}}
\newcommand{\drhopm}{\delta\rho^{\pEN}}
\newcommand{\drhomp}{\delta\rho^{\pNE}}
\newcommand{\hatrhop}{\hat{\rho}^{\pE}}
\newcommand{\hatrhom}{\hat{\rho}^{\pN}}
\newcommand{\hatrhopm}{\hat{\rho}^{\pEN}}
\newcommand{\hatdrhop}{\delta\hat{\rho}^{\pE}}
\newcommand{\hatdrhom}{\delta\hat{\rho}^{\pN}}
\newcommand{\hatdrhopm}{\delta\hat{\rho}^{\pEN}}
\newcommand{\tilderhop}{\tilde{\rho}^{\pE}}
\newcommand{\tilderhom}{\tilde{\rho}^{\pN}}
\newcommand{\tilderhopm}{\tilde{\rho}^{\pEN}}
\newcommand{\vp}{v^{\pE}}
\newcommand{\vm}{v^{\pN}}
\newcommand{\vpm}{v^{\pEN}}
\newcommand{\np}{n^{\pE}}
\newcommand{\nm}{n^{\pN}}
\newcommand{\npm}{n^{\pEN}}
\newcommand{\Jp}{J^{\pE}}
\newcommand{\Jm}{J^{\pN}}
\newcommand{\Jpm}{J^{\pEN}}

\newcommand{\alphap}{\alpha^{\pE}}
\newcommand{\alpham}{\alpha^{\pN}}

\newcommand{\sfRMm}{{\sf R}^M_m}
\newcommand{\JM}{J^M}

\newcommand{\alphac}{\alpha_{\rm c}}
\newcommand{\alphaM}{\alpha^{M}}
\newcommand{\alphaMm}{{\alpha^{M}_m}}
\newcommand{\alphaten}{\alpha^{10}}
\newcommand{\betaM}{\beta^{M}}
\newcommand{\betaMm}{{\beta^{M}_m}}
\newcommand{\sfR}{{\sf R}} 
\newcommand{\vR}{v_{\sf R}}
\newcommand{\Jf}{J_{\rm free}}
\newcommand{\vf}{v_{\rm free}}
\newcommand{\rhof}{\rho_{\rm free}}
\newcommand{\Jj}{J_{\rm jam}}
\newcommand{\vj}{v_{\rm jam}}
\newcommand{\rhoj}{\rho_{\rm jam}}

\newcommand{\la}{\langle}
\newcommand{\ra}{\rangle}
\newcommand{\beq}{\begin{equation}}
\newcommand{\eeq}{\end{equation}}
\newcommand{\bea}{\begin{eqnarray}}
\newcommand{\eea}{\end{eqnarray}}
\def\lsim{\:\raisebox{-0.5ex}{$\stackrel{\textstyle<}{\sim}$}\:}
\def\gsim{\:\raisebox{-0.5ex}{$\stackrel{\textstyle>}{\sim}$}\:}

\numberwithin{equation}{section}
\thispagestyle{empty}
\title{\Large  
{\bf Continuous and first-order
jamming transition in crossing pedestrian traffic flows}\\ 
\phantom{XXX}
}

\author{{H.J.~Hilhorst, J.~Cividini, and C.~Appert-Rolland}\\[5mm]
{\small Laboratoire de Physique Th\'eorique, b\^atiment 210}\\
{\small Universit\'e Paris-Sud and CNRS,
91405 Orsay Cedex, France}\\}

\maketitle

\begin{small}
\begin{abstract}
\noindent
After reviewing the main results obtained within a model for 
the intersection of
two perpendicular flows of pedestrians,
we present a new finding:  the changeover of the jamming transition
from continuous to first order when the size of the  
intersection area increases.
\vspace{2mm}

\end{abstract}
\end{small}
\vspace{12mm}


\newpage


\section{Introduction} 
\label{sect_introduction}

\begin{figure}[bht]
\begin{center}
\scalebox{.40}
{\includegraphics{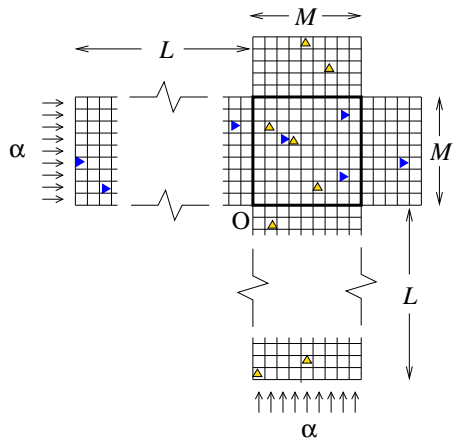}} 
\end{center}
\caption{\small Intersection of two one-way streets of width $M$.  
  The blue particles (\bluew{$\blacktriangleright$}) move eastward and
  the orange particles (\orangew{$\blacktriangle$})
  northward. The parameter $\alpha$ determines the particle injection rate.
  The region bordered by the heavy solid line is the `intersection
  square'. Figure taken from \cite{cividini_a_h2013a}.
} 
\label{fig_intersectinglanes}
\end{figure}

In this talk we will deal with crossing flows of 
pedestrians, modeled as hard core
particles that move on a lattice.
The geometry of interest to us is shown in
Fig.\,\ref{fig_intersectinglanes}: 
it represents two intersecting streets of width $M$
and, in principle, infinite length.
There are two kinds of particles, those moving east (blue) 
and those moving
north (orange or red); each move covers a single lattice distance.
When an eastbound 
and a northbound particle have the same target site,
there is a question of which one has priority. 
The answer is given by the rules of motion of the model, that is,
by the particle update algorithm.

Frequently used algorithms are random sequential
update, parallel update, alternating parallel update, sublattice
update, random shuffle update, and so on 
(some definitions are given in \cite{appert-rolland_c_h2011a}
and in \cite{rajewsky_s_s_s1998}). 
Recently we were looking for an algorithm that would (i)
make every particle advance, as long as it is not blocked, at unit speed,
and (ii) provide a natural answer to the priority question.
Wishing to avoid certain disadvantages of existing algorithms
we were led to introduce the {\it frozen shuffle update\,}
\cite{appert-rolland_c_h2011a,appert-rolland_c_h2011b}.
While operating in continuous time 

$\bullet$ this algorithm 
assigns to each particle $i$ injected into the system a 
{\it phase\,} $\tau_i\in[0,1]$ 
which is the fractional part of the time $t_i$ of injection;
the time intervals during which an injection
site is empty ({\it i.e.,} between the departure of an occupying
particle and the arrival of the next one)
are i.i.d.~exponential variables of average $1/a$ ; the injection
probability $\alpha=1-\ee^{-a}$
during a unit time interval is a convenient control parameter;

$\bullet$ during the $n$th unit time interval, $n \leq t <n+1$, 
this algorithm visits the particles in order of increasing
phases and advances particle $i$ at time $t=n+\tau_i$
at the condition that its target site be empty.

This talk is centered around the crossing street system with 
frozen shuffle update; 
we will nevertheless stress, where applicable, 
the robustness of our results under change of algorithm.


\section{One-dimensional model}

\begin{figure}
\begin{center}
\scalebox{.30}
{\includegraphics{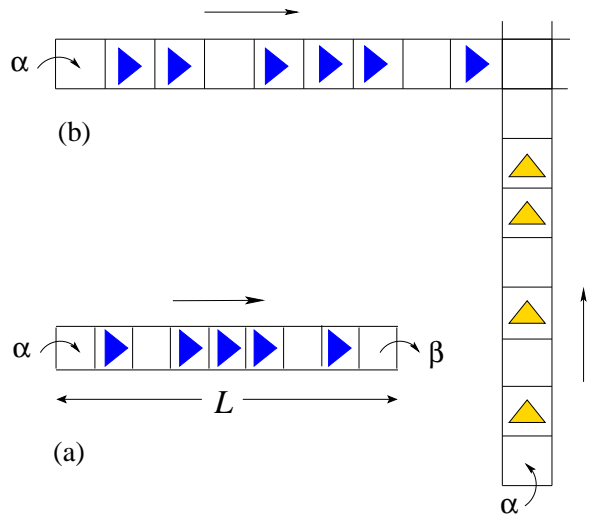}} 
\end{center}
\caption{\small (a) A single lane of finite length $L$
with an injection probability $\alpha$ and an exit probability
$\beta$.
(b) Case $M=1$ of Fig.\,\ref{fig_intersectinglanes}: two
intersecting lanes. The exit probability from the intersection site
is unity; however, their mutual obstruction leads for each lane
to an {\it effective\,} transit probability $\beta_1^1=\tfrac{1}{2}$
through the intersection site. 
} 
\label{fig_twocross}
\end{figure}

\begin{figure}
\begin{center}
\scalebox{.30}
{\includegraphics{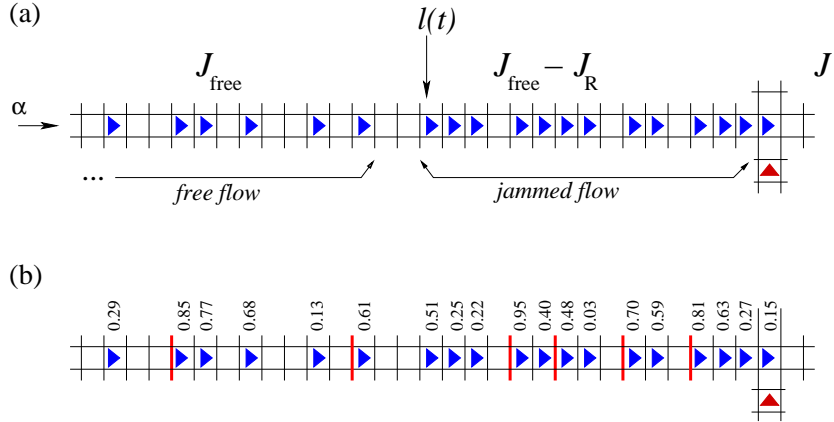}} 
\end{center}
\caption{\small Intersection of two single lanes.
(a) A waiting line of length $\ell(t)$ at the intersection site
divides the horizontal lane into a free flow and a jammed flow
domain.\,\, (b) The random phases of the particles in the horizontal
lane define a division of them
into platoons. A heavy vertical red bar has been placed to the
left of the last particle of each platoon. 
Those in the jammed flow domain are compact.
} 
\label{fig_waitingline}
\end{figure}

\begin{figure}
\begin{center}
\scalebox{.25}
{\includegraphics{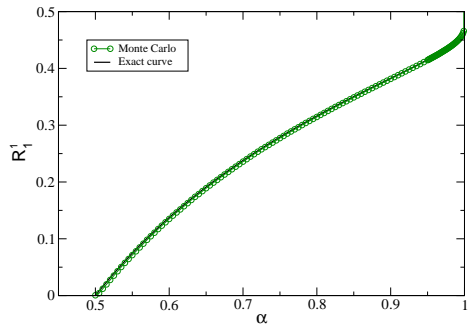}} 
\end{center}
\caption{\small Monte Carlo and analytic result (\ref{xM1})
for the reflection coefficient $\sfR=\sfR_1^1$ coefficient for $M=1$.
Figure adapted from \cite{hilhorst_a2012}.
} 
\label{fig_R1}
\end{figure}

On the one-dimensional lattice 
of Fig.\,\ref{fig_twocross}a the model described above reduces to
a totally asymmetric simple exclusion process (TASEP).
We label the
sites $x=-L,-L+1,\ldots,-2,-1,$ the origin being chosen at the exit.
For an exit probability $\beta=1$,
due to the specific way \cite{appert-rolland_c_h2011b}
of injecting the particles,
none of them ever blocks its successor;
all particles will then traverse this lattice at unit speed, $\vf=1$,
and will be said to be in a state of `free flow'.
The statistics of the free flow is fully known. In particular, the
particle density $\rhof$ and the particle current $\Jf=\vf\rhof$ 
are given by 
\beq
\rhof(\alpha)=\Jf(\alpha) = \frac{a}{1+a}\,.
\label{xJfreealpha}
\eeq
For an exit probability
$\beta<1$, whenever a particle wishes to move off the last site
and thereby exit the system, this step is executed only with
probability $\beta$. If refused, the particle stays where it
is and waits a unit time interval until it can make its next attempt. 
The blocking of the exit may lead to
blockings further down the lane and may create an
intermittent or permanent waiting line.

A particle undergoing a blocking by its predecessor will be said to
belong from that moment on to the {\it waiting line.} Let us denote 
the site of the leftmost particle in the waiting line by $x=-\ell(t)$,
where $\ell(t)=1,2,3,\ldots,L$;
we set $\ell(t)=0$ if none of the particles in the system has
ever been blocked.
The  site $x=-\ell(t)$ separates a `free flow domain' to its left
from a `jammed domain' to its right; we will therefore say
that it is the location of a {\it domain wall.}  
In the moving frame that has the domain wall as its origin, the
density profile to the left is strictly constant and equal to $\rhof$,
whereas to the right it decays after some weak oscillations
rapidly to a higher value $\rhoj$.
A study of this profile was carried out in Ref.\,\cite{appert-rolland_c_h2011b} 
and has, incidentally, raised an interesting 
question  \cite{cividini_h_a2013c} of the validity of domain
wall theory \cite{kolomeisky98,pigorsch_s00,santen_a02} for this problem. 

The domain wall $\ell(t)$ performs a random walk. 
At fixed $\beta$ we expect that for low enough $\alpha$
it is localized within some
finite penetration depth $\xi(\alpha)$ from the exit 
(the system is in a state of free flow),
and that for high enough $\alpha$ 
it is localized within $\xi(\alpha)$ from the entrance
(the system is jammed).   
In the vicinity of a critical value $\alpha=\alphac(\beta)$
the penetration depth becomes of order $L$ and
in the limit $L\to\infty$ 
a sharp critical point $\alphac(\beta)$ arises on the $\alpha$ axis.

Whereas the particle density $\rhof$ and the current $\Jf$ in the free
flow domain are known, the analogous quantities $\rhoj$ and $\Jj$ 
in the jammed domain have to be calculated.
The key to the exact solution resides in the concept of a 
{\it platoon}, defined as 
a maximal sequence of successive particles having increasing phases
(see Fig.\,\ref{fig_waitingline}b).
This concept is therefore linked to the frozen shuffle update
algorithm. A platoon is said to be compact%
\footnote{In our earlier work we reserved the name `platoon' for
what we now call `compact platoons'.}
if, at any integer instant
of time $t=n$, there are no empty sites between its constituent
particles.
An elementary calculation {\cite{appert-rolland_c_h2011b}
shows that the average number $\nu$ of 
particles in a platoon is given by
\beq
\frac{1}{\nu} =  1 + \frac{1}{a} - \frac{1}{\alpha}.
\label{xnu}
\eeq
The argument leading to $\rhoj$ and $\Jj$ begins by 
considering
the exit at $x=0$ in contact with the jammed domain.
It may be shown \cite{appert-rolland_c_h2011b} that in the jammed domain  

$\bullet$ each platoon is compact;

$\bullet$ two platoons are separated by either 0 or 1 empty site and
the density of the empty sites is $\beta/(\nu+\beta)=1-\rhoj$.

Both features are illustrated in Fig.\,\ref{fig_waitingline}b. 
Employing these properties we deduce that
$\Jj=\beta\rhoj$ with the current $\Jj$ 
given by the equation
\beq
\frac{1}{\Jj}-\frac{1}{\Jf} =
  \frac{1}{\beta}-\frac{1}{\alpha}\,. 
\label{xJjam}
\eeq
The current $J$ effectively passing through the system 
is $J=\min(\Jf,\Jj)$ and hence} Eq.\,(\ref{xJjam})
shows that the critical jamming point occurs for
\beq
\alphac=\beta.
\label{xalphac}
\eeq
Although Eqs.\,(\ref{xJjam}) and (\ref{xalphac}) are elegant and simple,  
we have found no quick way to see that they must be true:
the system does not have
the particle-hole symmetry that facilitates the analysis of certain
other TASEPs.

A final remark is that, unlike with other
updates, the current $\Jj$ here depends not only on $\beta$ but also on
$\alpha$, which comes in through the platoon structure.
For the intersecting streets to be studied below we have $\beta=1$,
but $\Jj$ will continue to depend on $\alpha$.


\section{Two crossing lanes: the case $M=1$}

We now consider the crossing of the two
single lanes shown in Fig.\,\ref{fig_twocross}b, where the exit
probability from the intersection site is unity; this is
Fig.\,\ref{fig_intersectinglanes} for the special case $M=1$.
By an `exact solution' we will mean an exact expression 
[such as (\ref{xJjam}) and (\ref{xalphac})]
for the critical point $\alphac$ and
current $J$ as a function of the injection probability $\alpha$. 
We expect again that there is a critical value $\alpha=\alphac$
such that $J(\alpha)=\Jf(\alpha)$ 
for $\alpha<\alphac$ and
\bea
J(\alpha) &=& \Jj(\alpha) \nonumber\\[2mm]
          &\equiv& 
[1-\sfR(\alpha)]\Jf(\alpha),  \qquad  \alpha>\alphac\,,
\label{defR}
\eea
and the challenge is to
calculate $\alphac$ and $J_{\rm jam}(\alpha)$ in the jammed phase.
The second line of (\ref{defR}) is a rewriting 
which shows that $\Jf$ and $\sfR\Jf$ are analogous to
an incident and a reflected wave, respectively
and defines the {\it reflection coefficient\,} $\sfR$; 
in Fig.\,\ref{fig_waitingline}a the reflected wave has advanced to the point
marked $\ell(t)$. 
Since $\sfR(\alpha)$
vanishes for $\alpha<\alphac$ axis and varies between $0$ and $1$
for $\alpha>\alphac$, we may employ it as the order
parameter of the transition.

Fig.\,\ref{fig_R1} shows the behavior of $\sfR(\alpha)$ 
for the system of two crossing
lanes of Fig.\,\ref{fig_twocross}.
This case is in fact exactly soluble \cite{appert-rolland_c_h2011c}.
It leads to $\alphac=\tfrac{1}{2}$ and
\beq
\sfR(\alpha) = \frac{\nu}{2\nu+1}\,\frac{2\alpha-1}{\alpha}\,, \qquad
\alpha>\alphac=\tfrac{1}{2}\,. 
\label{xM1}
\eeq
Comparison of this value $\alphac=\tfrac{1}{2}$ 
to Eq.\,(\ref{xalphac}) shows that it is as if each lane
exerts a blocking effect on the other one equivalent to an effective
exit probability that we will call $\beta_1^1$ (for a reason to become
clear) and that takes the value $\beta_1^1=\tfrac{1}{2}$.
The curve (\ref{xM1}) is shown in Fig.\,\ref{fig_R1}.

It is worthwhile to note that in the jammed phase the front of the
reflected ``wave'' propagates at a constant average speed,
\beq
\langle\ell(t)\rangle = \vR t, \qquad \alpha>\alphac\,,
\eeq
in which $\langle\ldots\rangle$ is an ensemble average and
where $\vR(\alpha)$ and $\sfR(\alpha)$ are related by
\cite{hilhorst_a2012}
\beq
\vR = \frac{\alpha\nu\sfR}{\alpha\sfR+(1-\alpha)\nu}\,.
\label{relvRsfR}
\eeq

To conclude we remark
that the same system of two crossing lanes can still be solved exactly 
\cite{appert-rolland_c_h2011c} for
unequal entrance probabilities $\alpha_1$ and $\alpha_2$
and exit probabilities $\beta_1$ and $\beta_2$ (real ones, not the
effective one mentioned above)  
less than unity.


\section{Streets of width $M>1$}

\subsection{Theory}

For crossing streets of width $M>1$ we have not so far found any exact
solutions. The division of the sequence of particle into platoons
seems no longer helpful.
We do not exclude that somebody can find the solution for a
$2\times 2$ intersection square, or perhaps for the asymmetric case
of an $M\times 1$ square. 
However, if the general case seems out of reach 
theoretically, accurate numerical studies are possible that reveal 
  interesting properties. We prepare the ground by defining the order
  parameters that will be relevant.

We will restrict ourselves to two crossing streets of the same width
$M$. We will
number the lanes from the inner ones outward by an index
$m=1,2,\ldots,M$.
To each $m$ there corresponds an eastward and a northward lane and by
symmetry the two are statistically identical (we never found any
hint of symmetry breaking).
In the $m$th lane we now have a relation analogous to (\ref{defR}) but
augmented with the indices $M$ and $m$
indicating the street width and the lane number,
\beq
J_m^M(\alpha)=[1-R^M_m(\alpha)]J_{\rm free}(\alpha),
\label{defRMm}
\eeq
and a similar generalization of Eq.\,(\ref{relvRsfR}).
The $R^M_m$ are now the order parameters. 
We may formally write the reflection coefficients as \cite{hilhorst_a2012}
\beq
\sfRMm = \frac{\nu\betaMm}{\nu+\betaMm}
\left( \frac{1}{\betaMm}-\frac{1}{\alpha} \right).
\label{xsfR}
\eeq
which for $M=m=1$ and $\beta_1^1=\tfrac{1}{2}$ reduces to (\ref{xM1}),
and where the $\betaMm$ now has the interpretation of 
an effective exit probability from the $m$th lane.
However, for $M\geq 2$ we can find the $\sfRMm$ 
(or equivalently the $\betaMm$) only by simulation.


\subsection{Simulation: memory boundary conditions}
\label{secmbc}

Simulations are necessarily carried out on finite lattices.
A finite value of $L$ (see Fig.\,\ref{fig_intersectinglanes})
sets an upper limit to the length $\ell(t)$ of the waiting line 
and will cause a rounding of the transition.
We have conceived \cite{hilhorst_a2012} an algorithm 
allowing the simulation of
systems having $L=\infty$, using only a finite number of variables.
The trick is to consider the $M\times M$ interaction square with
special boundary conditions, termed `memory boundary
conditions'. For each lane we keep track of the particle
positions {\it in\,} the intersection square plus one extra variable
which, essentially, is the length of the waiting line in that lane.
This eliminates all finite length effects 
and the sharpness of the transition point increases with the
duration of the simulation. 


\subsection{Jamming for street widths $M\lesssim 24$}

\begin{figure}
\begin{center}
\scalebox{.35}
{\includegraphics{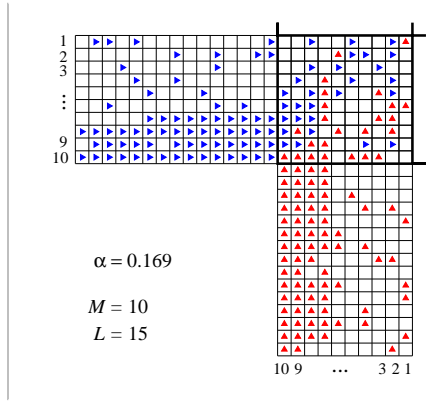}} 
\end{center}
\caption{\small Snapshot of intersecting streets of width $M=10$.
The inner lanes are jammed and the outer ones in a state of free
flow. 
Figure taken from \cite{hilhorst_a2012}.
} 
\label{fig_cross10}
\end{figure}

\begin{figure}
\begin{center}
\scalebox{.25}
{\includegraphics{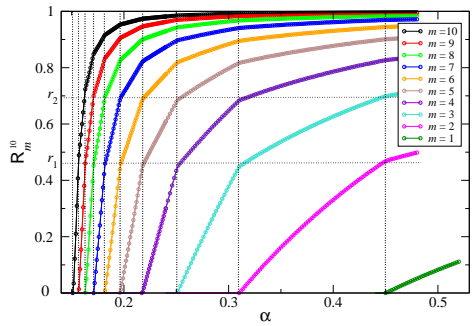}} 
\end{center}
\caption{\small Reflection coefficients for $M=10$.
Figure taken from \cite{hilhorst_a2012}.
} 
\label{fig_R10}
\end{figure}

We carried out simulations
of the intersecting streets of widths up to $M=24$.
It appears \cite{hilhorst_a2012}, 
at least for the values $M\lesssim 24$ that were investigated,
that the $m$th lane undergoes jamming at 
a critical point $\alpha^M_m$ and that 
$\alpha^M_M<\alpha^M_{M-1}<\ldots<\alpha^M_1$.
We call $\alpha^M_M$ the {\it principal critical point.}
Fig.\,\ref{fig_cross10}, obtained with the traditional finite
$L$ simulation algorithm, shows a snapshot of the system with
$M=10$ and $L=15$ which has its inner lanes jammed and its outer ones in
a state of free flow. 
Fig.\,\ref{fig_R10} shows the ten reflection coefficients
$\sfRMm(\alpha)$ for the $M=10$ system,
which by (\ref{defRMm}) are directly equivalent to
the currents $\JM_m(\alpha)$
[this one and all further simulations were carried out with the memory boundary
conditions of section \ref{secmbc}, {\it i.e.,} for $L=\infty$].

\begin{figure}
\begin{center}
\scalebox{.25}
{\includegraphics{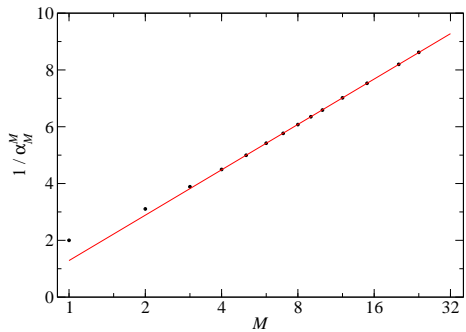}} 
\end{center}
\caption{\small Principal critical point as a function of $M$.
Figure taken from \cite{hilhorst_a2012}.
} 
\label{fig_alphaMM}
\end{figure}

It appears that the principal critical point $\alphaM_M$ decreases with $M$
and that its behavior is very well approximated by
\beq
\alphaM_M \simeq \frac{1}{A+B\log M}\,, \qquad \mbox{$M \gsim 4$},
\label{xalphaMM}
\eeq 
as shown in Fig.\,\ref{fig_alphaMM}, where $A=1.287$ and $B=2.306$.
The high precision of these results 
is due to the elimination of finite length effects.

\subsection{Jamming for larger street widths}

\begin{figure}
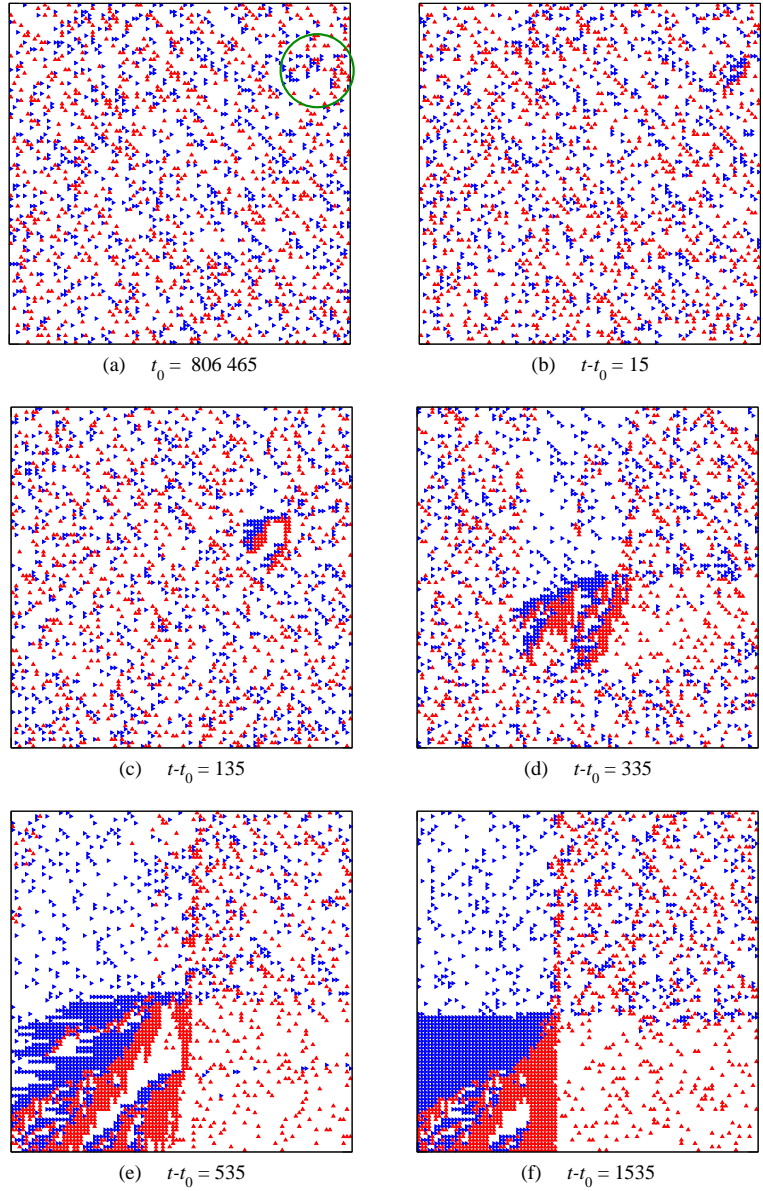

\begin{center}
\scalebox{.30}
{\includegraphics{Figpersp8a.eps}}\qquad
\scalebox{.30}
{\includegraphics{Figpersp8b.eps}}\\[3mm] 
\scalebox{.30}
{\includegraphics{Figpersp8c.eps}}\qquad
\scalebox{.30}
{\includegraphics{Figpersp8d.eps}}\\[3mm]
\scalebox{.30}
{\includegraphics{Figpersp8e.eps}}\qquad 
\scalebox{.30}
{\includegraphics{Figpersp8f.eps}} 
\end{center}
\caption{\small Snapshots of the nucleation instability in a free flow
  state (see text).
} 
\label{fig_snapshot0}
\end{figure}

The question of the {$M\to\infty$} 
limit of the critical point was
asked also in the context of the BML model
\cite{biham_m_l1992,ding_j_w2011}, but none of the
authors has been able to state whether or not
this point goes to zero in that limit.
On the basis of our above results
one might guess that (\ref{xalphaMM}) is the correct asymptotic law 
and hence that for the model of this work
the principal critical point does tend to zero with increasing $M$.
However, when the simulations are pushed to larger lattice sizes, a
novelty appears \cite{hilhorst_cividini_a2013}. 
This new phenomenon is already suggested by the fact that the initial slope of
$R_M^M(\alpha)$ is of increasing steepness as $M$ becomes larger,
and that infinite steepness would correspond to a first order
transition. Hence for growing $M$ the transition seems on its
way of becoming first-order.
This is confirmed by further investigation.
In fact, for large enough $M$ and when $\alpha$ increases,
the free flow phase appears to become metastable: 
by means of a nucleation mechanism 
it may  irreversibly turn into a jammed phase.
This is exemplified in Fig.\,\ref{fig_snapshot0}.
A system of linear size $M=100$ is started at time $t=0$ in a free
flow configuration with $\alpha=0.08729$, both free flows having just
arrived at the entrance of the empty interaction square. 
After a transient of no more than a few hundred time steps the system
settles in what seems to be a stationary state in which it stays for 
the first 800\,000 time steps.  
Fig.\,\ref{fig_snapshot0}a shows the particle configuration on the
intersection square at a certain time $t_0=806\,465$. 
It is characteristic of the stationary state; the particle
configuration shows at certain points in space 
small fluctuating densifications which
normally appear and disappear. However, the one in the circle
starts acting as the nucleus of a jammed domain. 
Figs.\,\ref{fig_snapshot0}b-\ref{fig_snapshot0}f show 
this domain at a succession of later times.
While growing from the south and the west, it evaporates 
(but less fast) on the north and east side, with as a net result a
displacement towards the south-west corner of the interaction square.
Once it arrives there, it sticks to the entrance sites and blocks a set of
horizontal and vertical inner lanes. 
We recall that the simulation includes the memory
variables representing the lengths of the waiting line in each lane,
even though these lines are not shown in the figures.
In the course of time, the exact 
set of inner lanes that are blocked is subject to fluctuation
but the jammed domain remains stable:
we have continued the simulation until time $t=11\times 10^6$
without seeing it disappear.
The final jammed state resembles closely a free flow state with an
effective street width having a reduced value $M^\prime<M$. 

The conclusion is that the continuous jamming transition observed in
earlier work for $M\lesssim 24$ turns first order
when $M$ becomes larger. 
We have not been able to define a precise tricritical point 
$(\alphac,M_{\rm c})$ at which the changeover takes place, but
closer study, not reported here, reveals that the first order nature
begins to set in as soon as $M \gtrsim 25$.
As a consequence, the straight line of data points in
Fig.\,\ref{fig_alphaMM} cannot be continued beyond the range for which
it is shown. 

\section{Pattern formation and chevron effect}

\begin{figure}
\begin{center}
\scalebox{.30}
{\includegraphics{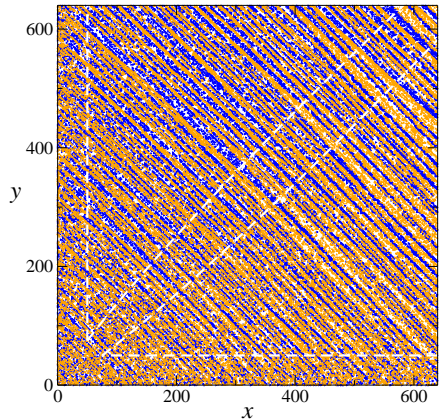}} 
\end{center}
\caption{\small Snapshot of an intersection square of linear size
$M=640$. The particles show stripe formation and a chevron effect (see
text). Figure taken from \cite{cividini_a_h2013a}.
} 
\label{fig_chevrons}
\end{figure}

Apart from the nucleation instability that 
it illustrates, Fig.\,\ref{fig_snapshot0}
still shows an altogether different phenomenon that is of interest.
It is the fact that in the free flow regime
the particles of the two types organize into
alternating diagonal stripes 
\cite{cividini_a_h2013a,cividini_h_a2013b,cividini_a_h2013}. 
Such stripes have indeed been observed in experiments and are also
reproduced by realistic `agent-based' models 
\cite{hoogendoorn_d2003,hoogendoorn_b2003,yamamoto_o2011}. 
In order to study them we
replaced the particle dynamics with mean field equations 
in which on every lattice site $\br$
two continuous variables, $\rhop(\br,t)$ and $\rhom(\br,t)$, 
represent the densities of the eastward and northward traveling
particles, respectively. These densities are postulated to satisfy 
\bea
\rhop_{t+1}(\br) &=& [1 - \rhom_t(\br)]\rhop_t(\br-\bex)
                                  + \rhom_t(\br+\bex)\rhop_t(\br),
\nonumber\\[2mm]
\rhom_{t+1}(\br) &=& [1 - \rhop_t(\br)]\rhom_t(\br-\bey) 
                                  +
                                  \rhop_t(\br+\bey)\rhom_t(\br),
\label{mfeqns2d}
\eea
where $\bexy$ are basis vectors. As an auxiliary problem
we solved these equations on an intersection square with periodic
boundary conditions, and found that the solution
manifests the same
stripe formation instability \cite{cividini_a_h2013a}
as observed in the simulation of the true problem with open boundaries.
We pointed out, however, that in the case of open boundaries
the stripes in fact are not exactly at
$45^\circ$ with respect to the main axes, but that they form chevrons
with a very weak opening angle, of the order of the degree. 
The effect appears again both in the particle simulation and in the
numerical solution of the mean field equations (\ref{mfeqns2d}) and
may be observed in Fig.\,\ref{fig_chevrons}. 
The slope of the stripes has two roughly constant, but distinct,
values the two triangular regions delimited by the dashed white lines.
The chevron angle may be determined accurately by sufficient
statistical averaging; it appears to be linear in $\alpha$.

In Ref.\,\cite{cividini_a_h2013a} we explained the origin of this
`chevron effect'. There is much to say about this phenomenon and
a detailed account is in preparation \cite{cividini_h_a2013b}; 
it includes cases with unequal
injection rates in the two perpendicular directions,
and with the intersection square subjected to cylindrical
boundary conditions. 
In Ref.\,\cite{cividini_a_h2013} we show how a particle may be
localized by the wake of another particle of the same type, a
mechanism which explains how global patterns are produced at the
microscopic scale.

\section{Outlook}

Simple stylized models like the present one are certainly not
meant to compete with more realistic ones 
\cite{hoogendoorn_b2003,yamamoto_o2011}.
The purpose of the model studied here is complementary. Since it may
be analyzed more fully, it sets a standard scenario with respect to
which others models may be discussed.
Models of this type may also draw our attention to phenomena in
traffic problems (of which the chevron effect is an example)
that remain easily hidden in more elaborate many-parameter models.
\vspace{1mm}

Real pedestrians in crossing flows have only two strategies available to avoid
collisions, and experimental observation shows that they use both.
The first one is adapting their speed; this is the
strategy implemented in the
present model. The second one is deviating their
trajectory; this would correspond to introducing the possibility of
sideways motion.
From observation we know that both strategies are used, but
that large deviations from straight trajectories are relatively rare.
We therefore consider the model presented here as a relevant starting
point. Incorporating lateral motion is left for later work. 
\vspace{1mm}

Certain properties of the present model may not survive the
introduction of sideways steps. In particular, the distinction between
$M$ successive jamming transitions in individual lanes is likely to
get blurred; and if so, the question of the nature of the jamming transition
will have to be asked anew. Other properties of this model, 
however, may well turn out to
be robust. One example is the predominance of the flow through the outer
lanes over those through the inner ones. Another one is the chevron effect;
we believe that in future observations and experiments
it will be worth looking for this effect.

\bibliographystyle{elsart-num}

\end{document}